\documentclass[aps,pra,twocolumn,showpacs,amsmath,amssymb]{revtex4}

\pdfoutput=1

\usepackage{graphicx}
\usepackage{dcolumn}
\usepackage{bm}

\begin{document}

\title{Precision measurement of the lifetime of the $6p \mbox{ } {}^{2}P_{1/2}$ level of Yb${}^{+}$}

\author{S. Olmschenk$^{1}$}
	\email{smolms@umd.edu}
\author{D. Hayes$^{1}$}
\author{D. N. Matsukevich$^{1}$}
\author{P. Maunz$^{1}$}
\author{D. L. Moehring$^{2}$}
\author{K. C. Younge$^{3}$}
\author{C. Monroe$^{1}$}
	\affiliation{${}^{1}$Joint Quantum Institute, University of Maryland Department of Physics and National Institute of Standards and Technology, College Park, MD 20742, USA \\
	${}^{2}$Max-Planck-Institut f\"{u}r Quantenoptik, 85748 Garching, Germany \\
	${}^{3}$FOCUS and Department of Physics, University of Michigan, Ann Arbor, Michigan, 48109, USA}

\date{\today}

\begin{abstract}
We present a precise measurement of the lifetime of the $6p \mbox{ } {}^{2}P_{1/2}$ excited state of a single trapped ytterbium ion (Yb${}^{+}$).  A time-correlated single-photon counting technique is used, where ultrafast pulses excite the ion and the emitted photons are coupled into a single-mode optical fiber.  By performing the measurement on a single atom with fast excitation and excellent spatial filtering, we are able to eliminate common systematics.  The lifetime of the $6p \mbox{ } {}^{2}P_{1/2}$ state is measured to be $8.12 \pm 0.02$ ns.
\end{abstract}

\maketitle

The ytterbium atomic system is noted for a wide range of applications, including the precise measurement of fundamental symmetries~\cite{demille:yb_pnc, kimball:pnc}, implementation of atomic frequency standards~\cite{fisk:171s12, blythe:171f, porsev:yb_lattice_clock}, generation of Bose and Fermi degenerate gases~\cite{takasu:yb_bec, fukuhara:yb_fermi}, and quantum information and computation~\cite{balzer:ybqip, olmschenk:statedetect, hayes:ultracold_exchange, gorshkov:alkaline_register}.  Atomic structure calculations are particularly challenging for ytterbium because of the complexity of its electronic configurations, making precision measurements of this system also valuable for tests of \textit{ab initio} theory~\cite{cowan:spectra, biemont:YbRydberg, safronova:yb-like}.

\begin{figure}
\centering
\includegraphics[width=1.0\columnwidth,keepaspectratio]{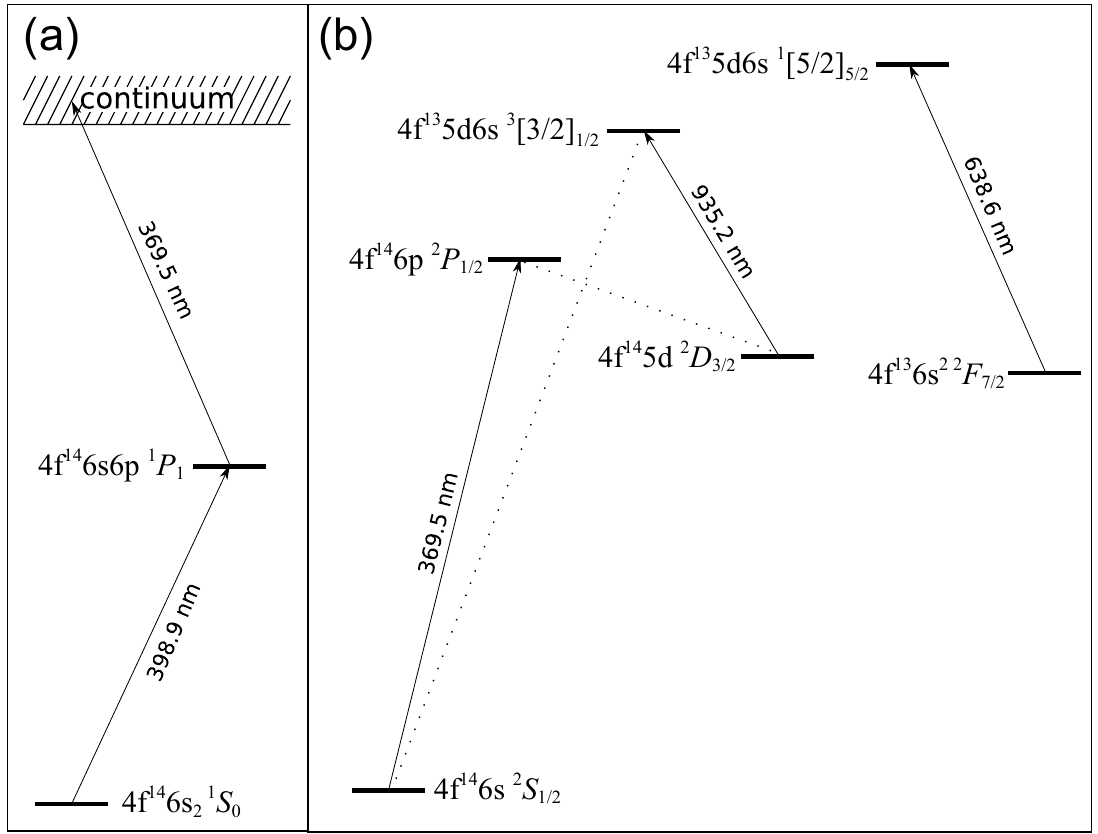}
\caption{(a) The partial energy diagram of the neutral Yb atom depicting photoionization.  (b) The relevant energy levels and transition wavelengths of Yb${}^+$.}
\label{fig:energy-level}
\end{figure}

A wide variety of experimental methods have been employed in an effort to obtain an accurate measurement of the excited states of the ytterbium ion (Yb${}^{+}$).  Methods for dipole-allowed transitions include delayed coincidence~\cite{burshtein:delayed-coin, blagoev:delayed-coin}, beam-foil~\cite{andersen:rare-earths}, Hanle effect~\cite{andersen:rare-earths, rambow:hanle}, laser-induced fluorescence of sputtered metal vapor~\cite{lowe:radiative} and plasma~\cite{li:YbIIlifetime}, beam-laser~\cite{berends:beam-laser-YbII, pinnington:lifetime}, and quantum jumps in single ions~\cite{taylor:d52, yu:lifetime}.  In this article, we report a measurement of the $6p \mbox{ } {}^{2}P_{1/2}$ excited state of the Yb${}^{+}$ atom using a time-correlated single-photon counting technique~\cite{young:cs-single-photon, moehring:lifetime}.  By using ultrafast pulses to excite a single trapped Yb${}^{+}$ atom and coupling the emitted photons into a fiber, we are able to eliminate many of the systematics present in earlier measurements.

Our experiment is performed on a single trapped ${}^{174}$Yb${}^+$ atom.  Ions are produced by first temporarily resistively heating a stainless steel tube filled with ytterbium (Yb) metal of natural isotopic abundance.  The result is an atomic beam perpendicular to and traversing two counterpropagating laser beams, one at 398.9 nm and the other at 369.5 nm.  Atoms are photoionized by a two-photon, dichroic, resonantly-assisted process as depicted in Fig.~\ref{fig:energy-level}(a)~\cite{olmschenk:statedetect}.

The resulting ion is confined in a radiofrequency (rf) linear Paul trap.  The trap consists of four parallel tungsten rods arranged symmetrically about and parallel to two opposing needles; the center-to-center spacing of adjacent rods is 1 mm, while the tip-to-tip spacing between the needles is about 2.6 mm.  An rf voltage at 37 MHz and amplitude of approximately 1 kV is applied to two of the four rods, providing transverse confinement of the ion.  To confine the ion in the axial direction, static voltages of about 80 V are applied to the needle electrodes.  The resulting secular frequencies are measured to be approximately 1 MHz, and the axial frequency of the trap is about 200 kHz.  Stray electric fields that result in micromotion by shifting the equilibrium position of the ion away from the rf node are compensated by applying small (order of 0.1 V) static potentials to the rod electrodes.  A single ion will typically remain in the trap for several weeks~\cite{olmschenk:statedetect}.

\begin{figure}
\centering
\includegraphics[width=1.0\columnwidth,keepaspectratio]{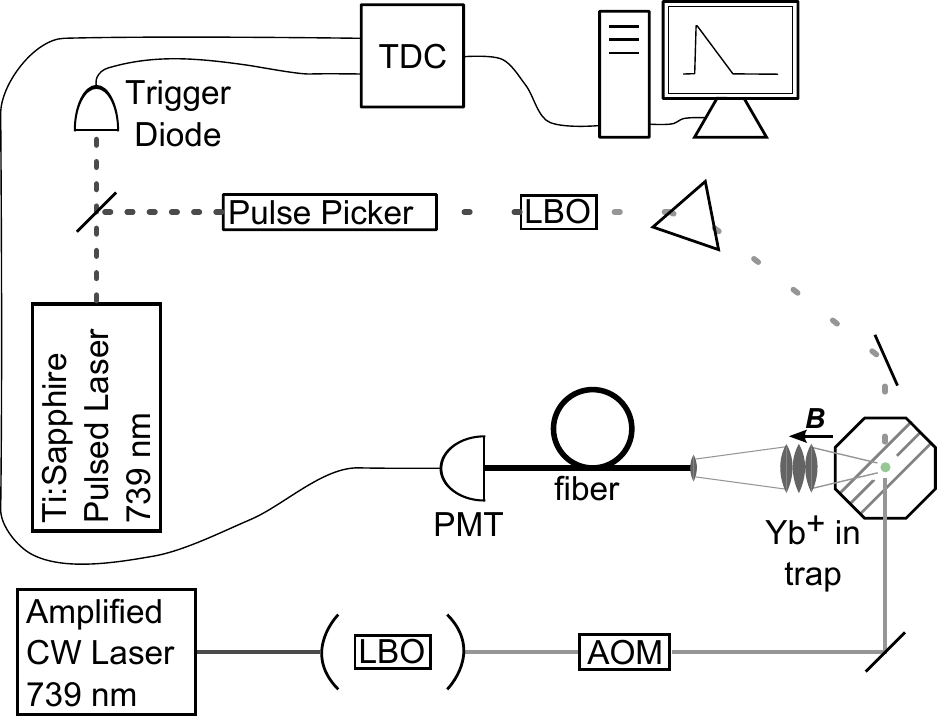}
\caption{The experimental setup used to determine the lifetime of the Yb${}^{+}$ $6p \mbox{ } {}^{2}P_{1/2}$ excited state.  The 935.2 nm and 638.6 nm lasers used for depopulating the metastable ${}^2D_{3/2}$ and ${}^2F_{7/2}$ states, respectively, are not shown.  LBO: lithium triborate nonlinear crystal; AOM: acousto-optic modulator; TDC: time-to-digital converter; PMT: photomultiplier tube; \textbf{\textit{B}}: applied magnetic field.}
\label{fig:expt-setup}
\end{figure}

The Yb${}^{+}$ atom is cooled by a four-level excitation scheme~\cite{bell:four-level}, as shown in Fig.~\ref{fig:energy-level}(b).  An amplified continuous wave (cw) diode laser at 739 nm is frequency doubled to 369.5 nm by a lithium triborate (LBO) nonlinear crystal in a cavity.  This light is detuned by about half a linewidth from the ${}^{2}S_{1/2} \leftrightarrow {}^{2}P_{1/2}$ transition of the Yb${}^{+}$ atom, providing effective Doppler cooling.  The ${}^{2}P_{1/2}$ state decays to the ${}^{2}D_{3/2}$ metastable state with a probability of about 0.005~\cite{olmschenk:statedetect}.  Another cw laser at 935.2 nm, resonant with the ${}^{2}D_{3/2} \leftrightarrow {}^{3}[3/2]_{1/2}$ transition, rapidly depletes the ${}^{2}D_{3/2}$ state, returning the ion to the cooling cycle.  A few times per hour, the ion is found in the low-lying, long-lived ${}^{2}F_{7/2}$ level, most likely as a result of collisions~\cite{lehmitz:pop-trap}.  We return the ion to the cooling cycle by using a 638.6 nm cw laser resonant with the ${}^{2}F_{7/2} \leftrightarrow {}^{1}[5/2]_{5/2}$ transition~\cite{taylor:d52}.

A block diagram of the experimental setup is shown in Fig.~\ref{fig:expt-setup}.  The cw lasers are used to Doppler cool the ion, as described above.  An actively mode-locked Ti:sapphire laser at 739 nm produces 1 ps pulses with a repetition rate of about 81 MHz.  The pulses are passed through an electro-optic pulse picker that has an average extinction ratio better than 100:1 in the infrared.  We reduce the effective pulse repetition rate to about 5.4 MHz by allowing only one in every fifteen pulses to pass.  A critically phase-matched LBO crystal is used to frequency double each pulse to 369.5 nm, and this ultraviolet light is separated from the residual infrared by a prism.  Frequency doubling increases the effective average extinction ratio to about $10^{4}$:1.

In the experiment, the ion is first Doppler cooled for 200 $\mu$s, and then the cw 369.5 nm light is blocked.  The electro-optic pulse picker is then gated open for 390 $\mu$s, allowing a train of pulses to pass, where each subsequent pulse is separated from the preceding by about 186 ns.  Leakage of each pulse (in the infrared) through a mirror strikes a trigger diode, which sends an electronic pulse (start pulse) to one channel of a time-to-digital converter (TDC) that has a resolution of 4 ps~\cite{note:picoharp}.  Each frequency-doubled laser pulse excites the ion from ${}^{2}S_{1/2}$ to ${}^{2}P_{1/2}$ with near unit probability, and the subsequent spontaneous decay of the excited state produces a single photon~\cite{maunz:interference}.  The photons emitted by the ion are collected by an imaging system with numerical aperature $\sim$0.3, coupled into a single-mode fiber, and detected by a photomultiplier tube (PMT).  The arrival time of the consequent electronic pulses (stop pulses) from the PMT are recorded by the second channel of the TDC.

\begin{figure}
\centering
\includegraphics[width=1.0\columnwidth,keepaspectratio]{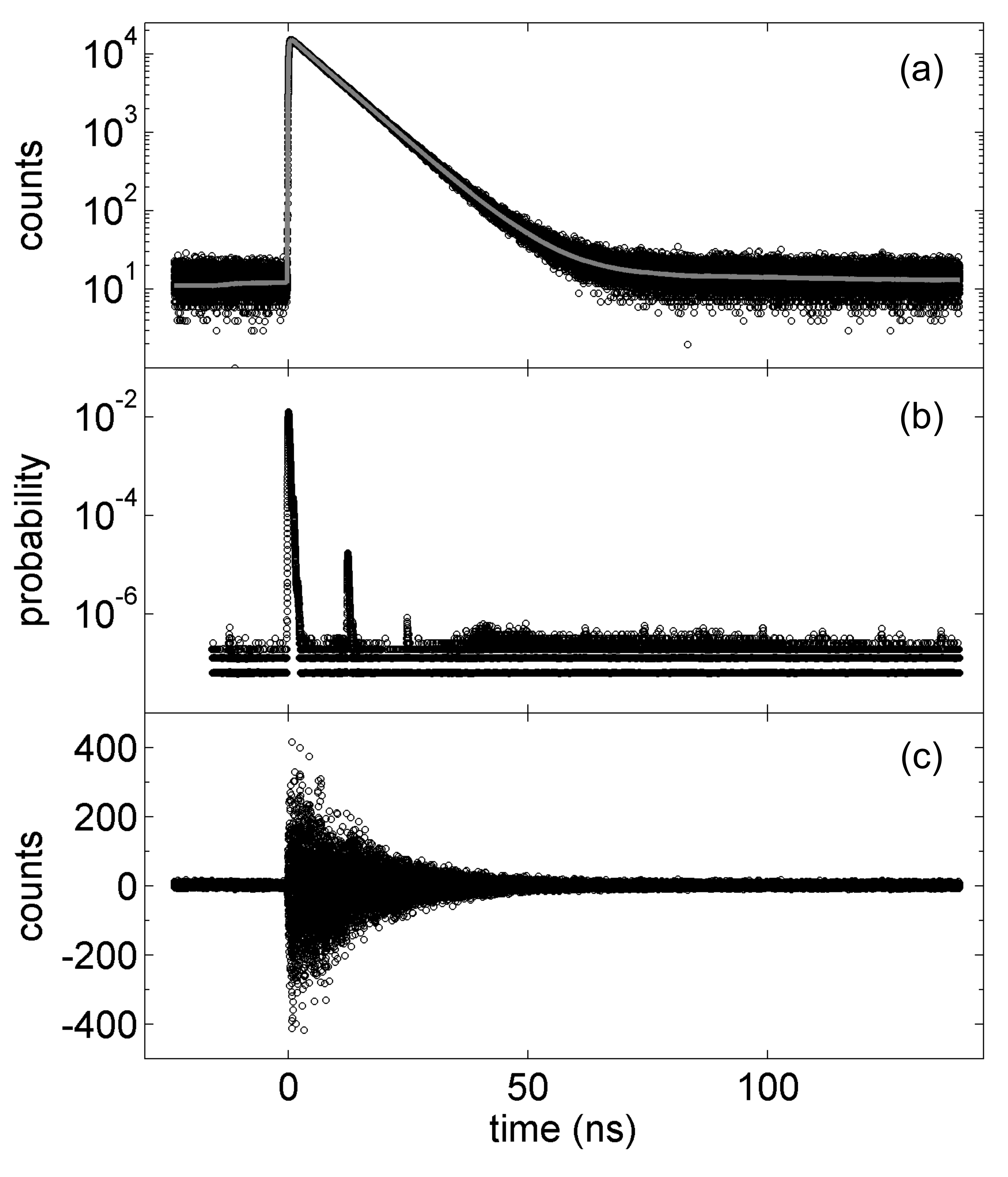}
\caption{Data used to determine the lifetime of the $6p \mbox{ } {}^{2}P_{1/2}$ excited state. (a) The data after correction for the differential nonlinearity of the TDC, with 4 ps binning, showing the number of photons (stop pulses) detected a given time following pulsed laser excitation (start pulses); the gray line is the fit, with functional form given by Eq.~\ref{eq:fit-function}.  (b) The normalized system response function.  Visible on the log-scale plot are the highly suppressed pulses from the mode-locked laser (average extinction ratio of about $10^{4}$:1), separated by the pulsed laser repetition rate of about 12.4 ns.  (c) Deviations of the data from the fitting function (residuals).}
\label{fig:lifetime-data}
\end{figure}

A histogram of the time difference between the arrivals of the two electronic pulses at the TDC displays the charateristic exponential decay of the excited atomic state (Fig.~\ref{fig:lifetime-data}(a)).  However, several other factors contribute to the overall shape of the histogram.  The pulse propagation times (of both the light and electronic pulses) results in an overall offset along the time axis.  Background counts result from PMT triggers due to either a background scattered photon or a dark count.  Scattered photons may be detected as a result of the laser pulse traversing the trap, producing a ``prompt peak'' in the data at the time of excitation.  Moreover, the finite response time of the detector dictates that the observed data is a convolution of all of these effects with the system response function.  Thus, the data shown in Fig.~\ref{fig:lifetime-data}(a) is described by the function
\begin{eqnarray}
	\label{eq:fit-function}
	F(t) & = & \sum_{t_n} \left[ A e^{-(t_n - t_0)/\tau} \Theta(t_n - t_0) \right. \nonumber \\
					&& \left. + B + C \delta_{n,0} \right] h(t - t_n) \nonumber \\
			 & = & B + C h(t - t_0) \nonumber \\
			 		&& + \sum_{t_n} A e^{-(t_n - t_0)/\tau} \Theta(t_n - t_0) h(t - t_n) ,
\end{eqnarray}
where $A$ is the amplitude of the exponential decay of the atomic state, $B$ is the background counts, $C$ is the integrated count contribution of the ``prompt peak,'' $h$ is the normalized system response function, $t_0$ is the time of excitation (used as a fit parameter, but shown in Fig.~\ref{fig:lifetime-data} as $t_0 = 0$), and $\Theta$ is the Heaviside step function.  The sum is over all possible time bins $t_n$ used in the histogram of the data.

We measure the system response function by coupling a small fraction of the light from the 369.5 nm pulse into an optical fiber, and directing the output onto the PMT.  Since the duration of the pulse (1 ps) is much shorter than the response time of the detector ($\sim$250 ps), the pulse is effectively a delta function input, allowing us to directly measure the system response function.  The measured system response function, $h$, is shown in Fig.~\ref{fig:lifetime-data}(b).  Also visible on the log-scale plot of this generalized system response function are the subsequent, highly suppressed pulses from the mode-locked laser (average extinction ratio of about $10^{4}$:1), separated by the approximately 12.4 ns repetition rate.

The differential nonlinearity of the TDC is characterized by directing light from an LED array to the fiber-coupler, and integrating this ``white noise'' input for several hours.  The differential nonlinearity of the TDC is found to result in stable, intermittent ripples with amplitude $< 6$\% peak-to-peak.  We use this measured signal (after normalization using a linear fit) to correct for the differential nonlinearity of the TDC by dividing the data (lifetime and system response measurements) by this signal.

The data analysis consists of computing Eq.~\ref{eq:fit-function} for a range of values for $A$, $B$, $C$, and $t_0$, and comparing this function to the data to determine $\chi^2$ for each combination of parameters.  The final fit ($A = 1.64 \times 10^4$, $B = 11.1$, $C = 6.79 \times 10^4$) is shown as the gray line in Fig.~\ref{fig:lifetime-data}(a), and the deviations of the data from this fit (residuals) are presented in Fig.~\ref{fig:lifetime-data}(c).  The statistical uncertainty in the fit is 0.002 ns.  The quoted uncertainty in the lifetime is dominated by systematic errors.

\begin{figure}
\centering
\includegraphics[width=1.0\columnwidth,keepaspectratio]{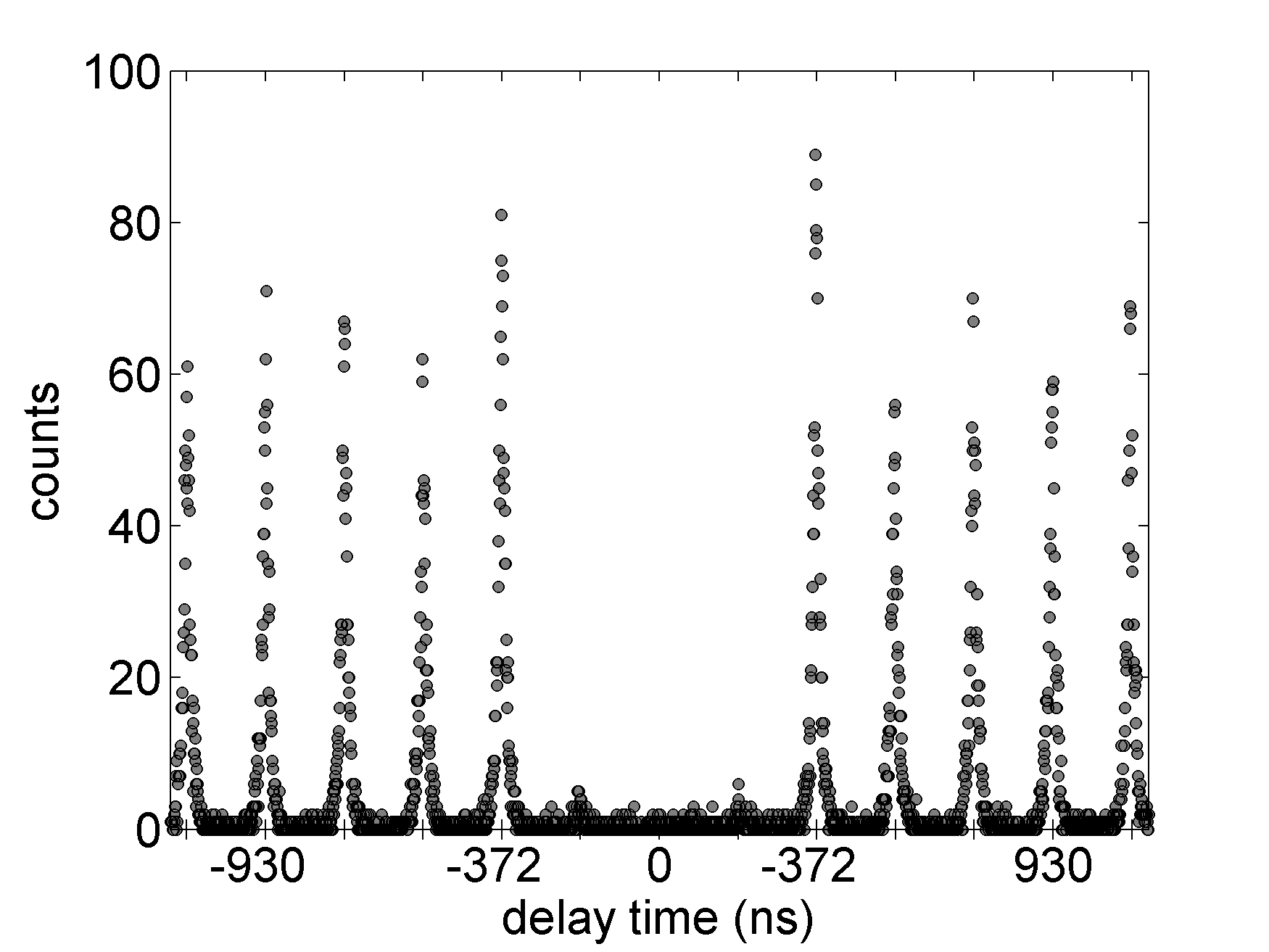}
\caption{Second-order correlation function used to determine the amplitude of quantum beats.  Through a combination of a quarter-wave plate and polarizers, we arrange the setup to observe $\sigma^{+}$-polarized light after a beamsplitter.  In using $\pi$-polarized pulses for excitation, two subsequent excitations should not both result in detection of a $\sigma^{+}$-polarized photon.  This is evidenced by the suppression of the first adjacent peaks in the second-order correlation function (at delay times of $\pm 186$ ns); the absence of a peak at zero delay time is evidence of an excellent source of single photons~\cite{maunz:interference}.}
\label{fig:three_missing_peak_g2}
\end{figure}

\begin{figure}
\centering
\includegraphics[width=1.0\columnwidth,keepaspectratio]{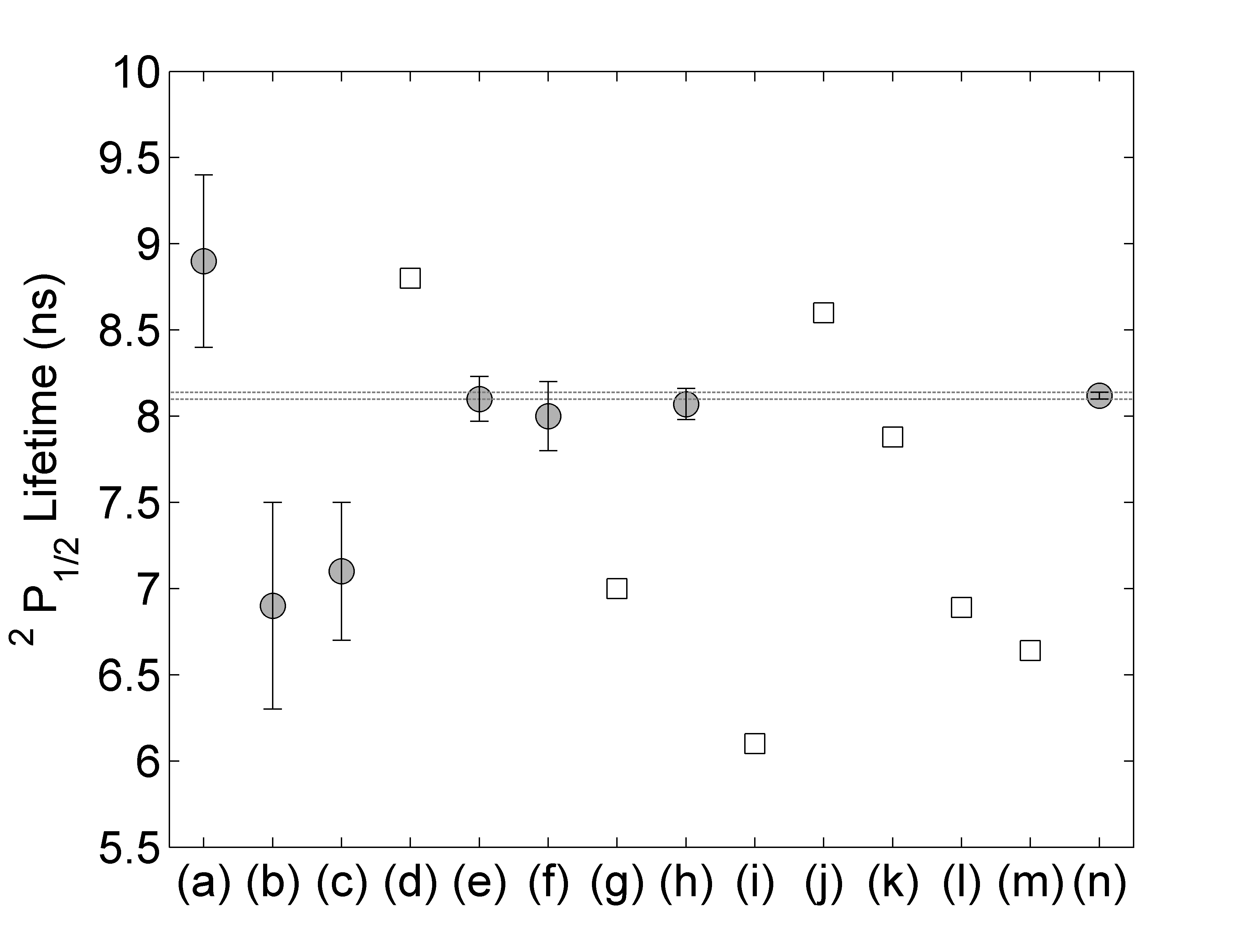}
\caption{Comparison of values obtained for $6p \mbox{ } {}^{2}P_{1/2}$ state lifetime.  (a) M. L. Burshtein et al. (1974), expt: delayed-coincidence~\cite{burshtein:delayed-coin}; (b) T. Andersen et al. (1975), expt: beam-foil~\cite{andersen:rare-earths}; (c) K. B. Blagoev et al. (1978), expt: delayed-coincidence~\cite{blagoev:delayed-coin}; (d) B. C. Fawcett et al. (1991), theory: Hartree-Fock~\cite{fawcett:lifetimes}; (e) R. W. Berends et al. (1993), expt: beam-laser~\cite{berends:beam-laser-YbII}; (f) R. M. Lowe et al. (1993), expt: laser-induced fluorescence~\cite{lowe:radiative}; (g) R. M. Lowe et al. (1993), theory: many-body perturbation theory~\cite{lowe:radiative}; (h) E. H. Pinnington et al. (1997), expt: beam-laser~\cite{pinnington:lifetime}; (i) E. H. Pinnington et al. (1997), theory: Coulomb approximation calculation~\cite{pinnington:lifetime}; (j) E. Bi\'{e}mont et al. (1998), theory: Hartree-Fock~\cite{biemont:Ybcalc}; (k) E. Bi\'{e}mont et al. (2002), theory: Hartree-Fock~\cite{biemont:YbRydberg}; (l) U. I. Safronova et al. (2009), theory: relativistic many-body perturbation theory~\cite{safronova:3rd_order_Yb_calc}; (m) U. I. Safronova et al. (2009), theory: relativistic many-body perturbation theory~\cite{safronova:3rd_order_Yb_calc}; (n) this work (gray dashed lines $\pm 1 \sigma$).}
\label{fig:compare-lifetime}
\end{figure}

Our experimental setup allows us to eliminate many possible systematics.  By using a single trapped atom, systematic errors such as pulse pileup, radiation trapping, subradiance and superradiance are eliminated~\cite{devoe:eomlifetime, devoe:superradiance, moehring:lifetime}.  Exciting this atom with an ultrafast pulse eliminates effects due to the application of light during the measurement interval, including background scattered light, multiple excitations, and ac Stark shifts~\cite{young:cs-single-photon, moehring:lifetime}.  Using the pulse picker to reduce the effective repetition rate of the ultrafast laser enables observation intervals much longer than the natural decay time of the excited state, allowing a fit of the data all the way to the tail end of the exponential where background events dominate.  Finally, by coupling the spontaneously emitted photons into a single-mode fiber, we nearly eliminate detection of background scattered light, including ``prompt peak'' photons scattered while the ultrafast pulse traverses the trapping region.

Two possible systematics that demand further investigation are hyperfine and Zeeman beats~\cite{haroche:qbeats}.  By using an even isotope of ytterbium (nuclear spin 0), we avoid the possibility of hyperfine beats.  On the other hand, at a magnetic field of 3.4 gauss the excited state Zeeman splitting is 3.1 MHz, so Zeeman beats could be significant.  However, $\pi$-polarized light is used to excite the ion, to prevent actively producing coherence between the Zeeman levels of the excited state.  In addition, by observing along the quantization axis, the radiation pattern of the atom suppresses the observation of $\pi$-polarized light so that only distinguishable decay channels are observed.  We are able to place an upper bound on the expected amplitude of the quantum beat signal by measuring a second-order correlation function of the spontaneously emitted photons, shown in Fig.~\ref{fig:three_missing_peak_g2}.

The setup for this correlation measurement is similar to Ref.~\cite{maunz:interference}, with two PMTs measuring the exit ports of a 50:50 beamsplitter.  For this measurement, a quarter-wave plate is inserted between the ion and the fiber-coupler, and polarizers are placed before the PMTs for detection of only $\sigma^{+}$-polarized light after the beamsplitter.  Given detection of a $\sigma^{+}$-polarized photon, since the excitation light is $\pi$-polarized, the immediately previous/subsequent spontaneously emitted photon ideally cannot be detected; this results in suppression of the first adjacent peaks of the correlation measurement (at delay times of $\pm 186$ ns), as seen in Fig.~\ref{fig:three_missing_peak_g2}.  The same errors that contribute to the generation of a quantum beat cause the first adjacent peaks of this second-order correlation function to be nonzero.  However, while in the second-order correlation signal the probabilities for coherence between the excited state Zeeman levels and detection of $\pi$-polarized photons add, in the quantum beat signal the amplitudes of these effects are multiplied~\cite{silverman:qbeats}.  The quantum beat amplitude would therefore be maximal if these errors contribute equally to the height of the first adjacent peaks in the second-order correlation signal, allowing us to place an upper bound on the modulation depth of the quantum beat signal of $< 0.03$.  Given this modulation depth, and a period of oscillation determined by the 3.1 MHz excited state Zeeman splitting, we determine quantum beats shift the measured lifetime of the excited state by less than $\pm 0.002$ ns.

The dominant systematic error was determined to be a residual ripple in the data due to either an uncompensated contribution of the differential nonlinearity of the TDC or incomplete characterization of the system response function.  By truncating the data prior to various points in the exponential decay/ripple (up to $t_0 + 30$ ns) and fitting the remaining data with a simplified version of Eq.~\ref{eq:fit-function}, we observe shifts in the fitted value of the lifetime.  We conservatively assign a systematic error large enough to encompass the fitted values from all of these truncated data sets, and arrive at a systematic error of $\pm 0.015$ ns.

The final value of the lifetime of the $6p \mbox{ } {}^{2}P_{1/2}$ state of Yb${}^{+}$ is measured to be $8.12 \pm 0.02$ ns.  In Fig.~\ref{fig:compare-lifetime}, our measurement is plotted alongside other experimental and theoretical values.  Our value agrees well with previous experimental measurements, with an improved uncertainty.  The clear disparity between our measurement and several of the theoretical results highlights the difficulty of the \textit{ab initio} calculations in this complex system, reinforcing the value of precision measurements in ytterbium to further an understanding of the structure of atoms.

\begin{acknowledgments}
We thank Anthea Monod for helpful discussions about the statistical analysis, and Marianna S. Safronova for discussions concerning atomic structure theory in ytterbium.  This work is supported by IARPA under ARO contract, the NSF PIF Program, and the NSF Physics Frontier Center at JQI.
\end{acknowledgments}


\begin{thebibliography}{37}
\expandafter\ifx\csname natexlab\endcsname\relax\def\natexlab#1{#1}\fi
\expandafter\ifx\csname bibnamefont\endcsname\relax
  \def\bibnamefont#1{#1}\fi
\expandafter\ifx\csname bibfnamefont\endcsname\relax
  \def\bibfnamefont#1{#1}\fi
\expandafter\ifx\csname citenamefont\endcsname\relax
  \def\citenamefont#1{#1}\fi
\expandafter\ifx\csname url\endcsname\relax
  \def\url#1{\texttt{#1}}\fi
\expandafter\ifx\csname urlprefix\endcsname\relax\def\urlprefix{URL }\fi
\providecommand{\bibinfo}[2]{#2}
\providecommand{\eprint}[2][]{\url{#2}}

\bibitem[{\citenamefont{DeMille}(1995)}]{demille:yb_pnc}
\bibinfo{author}{\bibfnamefont{D.}~\bibnamefont{DeMille}},
  \bibinfo{journal}{Phys. Rev. Lett.} \textbf{\bibinfo{volume}{74}},
  \bibinfo{pages}{4165} (\bibinfo{year}{1995}).

\bibitem[{\citenamefont{Kimball}(2001)}]{kimball:pnc}
\bibinfo{author}{\bibfnamefont{D.~F.} \bibnamefont{Kimball}},
  \bibinfo{journal}{Phys. Rev. A} \textbf{\bibinfo{volume}{63}},
  \bibinfo{pages}{052113} (\bibinfo{year}{2001}).

\bibitem[{\citenamefont{Fisk et~al.}(1997)\citenamefont{Fisk, Sellars, Lawn,
  and Coles}}]{fisk:171s12}
\bibinfo{author}{\bibfnamefont{P.~T.~H.} \bibnamefont{Fisk}},
  \bibinfo{author}{\bibfnamefont{M.~J.} \bibnamefont{Sellars}},
  \bibinfo{author}{\bibfnamefont{M.~A.} \bibnamefont{Lawn}}, \bibnamefont{and}
  \bibinfo{author}{\bibfnamefont{C.}~\bibnamefont{Coles}},
  \bibinfo{journal}{IEEE Trans. Ultrasonics, Ferroelectrics, and Frequency
  Control} \textbf{\bibinfo{volume}{44}}, \bibinfo{pages}{344}
  (\bibinfo{year}{1997}).

\bibitem[{\citenamefont{Blythe et~al.}(2003)\citenamefont{Blythe, Webster,
  Margolis, Lea, Huang, Choi, Rowley, Gill, and Windeler}}]{blythe:171f}
\bibinfo{author}{\bibfnamefont{P.~J.} \bibnamefont{Blythe}},
  \bibinfo{author}{\bibfnamefont{S.~A.} \bibnamefont{Webster}},
  \bibinfo{author}{\bibfnamefont{H.~S.} \bibnamefont{Margolis}},
  \bibinfo{author}{\bibfnamefont{S.~N.} \bibnamefont{Lea}},
  \bibinfo{author}{\bibfnamefont{G.}~\bibnamefont{Huang}},
  \bibinfo{author}{\bibfnamefont{S.-K.} \bibnamefont{Choi}},
  \bibinfo{author}{\bibfnamefont{W.~R.~C.} \bibnamefont{Rowley}},
  \bibinfo{author}{\bibfnamefont{P.}~\bibnamefont{Gill}}, \bibnamefont{and}
  \bibinfo{author}{\bibfnamefont{R.~S.} \bibnamefont{Windeler}},
  \bibinfo{journal}{Phys. Rev. A} \textbf{\bibinfo{volume}{67}},
  \bibinfo{pages}{020501(R)} (\bibinfo{year}{2003}).

\bibitem[{\citenamefont{Porsev et~al.}(2004)\citenamefont{Porsev, Derevianko,
  and Fortson}}]{porsev:yb_lattice_clock}
\bibinfo{author}{\bibfnamefont{S.~G.} \bibnamefont{Porsev}},
  \bibinfo{author}{\bibfnamefont{A.}~\bibnamefont{Derevianko}},
  \bibnamefont{and} \bibinfo{author}{\bibfnamefont{E.~N.}
  \bibnamefont{Fortson}}, \bibinfo{journal}{Phys. Rev. A}
  \textbf{\bibinfo{volume}{69}}, \bibinfo{pages}{021403(R)}
  (\bibinfo{year}{2004}).

\bibitem[{\citenamefont{Takasu et~al.}(2003)\citenamefont{Takasu, Maki, Komori,
  Takano, Honda, Kumakura, Yabuzaki, and Takahashi}}]{takasu:yb_bec}
\bibinfo{author}{\bibfnamefont{Y.}~\bibnamefont{Takasu}},
  \bibinfo{author}{\bibfnamefont{K.}~\bibnamefont{Maki}},
  \bibinfo{author}{\bibfnamefont{K.}~\bibnamefont{Komori}},
  \bibinfo{author}{\bibfnamefont{T.}~\bibnamefont{Takano}},
  \bibinfo{author}{\bibfnamefont{K.}~\bibnamefont{Honda}},
  \bibinfo{author}{\bibfnamefont{M.}~\bibnamefont{Kumakura}},
  \bibinfo{author}{\bibfnamefont{T.}~\bibnamefont{Yabuzaki}}, \bibnamefont{and}
  \bibinfo{author}{\bibfnamefont{Y.}~\bibnamefont{Takahashi}},
  \bibinfo{journal}{Phys. Rev. Lett.} \textbf{\bibinfo{volume}{91}},
  \bibinfo{pages}{040404} (\bibinfo{year}{2003}).

\bibitem[{\citenamefont{Fukuhara et~al.}(2007)\citenamefont{Fukuhara, Takasu,
  Kumakura, and Takahashi}}]{fukuhara:yb_fermi}
\bibinfo{author}{\bibfnamefont{T.}~\bibnamefont{Fukuhara}},
  \bibinfo{author}{\bibfnamefont{Y.}~\bibnamefont{Takasu}},
  \bibinfo{author}{\bibfnamefont{M.}~\bibnamefont{Kumakura}}, \bibnamefont{and}
  \bibinfo{author}{\bibfnamefont{Y.}~\bibnamefont{Takahashi}},
  \bibinfo{journal}{Phys. Rev. Lett.} \textbf{\bibinfo{volume}{98}},
  \bibinfo{pages}{030401} (\bibinfo{year}{2007}).

\bibitem[{\citenamefont{Balzer et~al.}(2006)\citenamefont{Balzer, Braun,
  Hannemann, Paape, Ettler, Neuhauser, and Wunderlich}}]{balzer:ybqip}
\bibinfo{author}{\bibfnamefont{C.}~\bibnamefont{Balzer}},
  \bibinfo{author}{\bibfnamefont{A.}~\bibnamefont{Braun}},
  \bibinfo{author}{\bibfnamefont{T.}~\bibnamefont{Hannemann}},
  \bibinfo{author}{\bibfnamefont{C.}~\bibnamefont{Paape}},
  \bibinfo{author}{\bibfnamefont{M.}~\bibnamefont{Ettler}},
  \bibinfo{author}{\bibfnamefont{W.}~\bibnamefont{Neuhauser}},
  \bibnamefont{and}
  \bibinfo{author}{\bibfnamefont{C.}~\bibnamefont{Wunderlich}},
  \bibinfo{journal}{Phys. Rev. A} \textbf{\bibinfo{volume}{73}},
  \bibinfo{pages}{041407(R)} (\bibinfo{year}{2006}).

\bibitem[{\citenamefont{Olmschenk et~al.}(2007)\citenamefont{Olmschenk, Younge,
  Moehring, Matsukevich, Maunz, and Monroe}}]{olmschenk:statedetect}
\bibinfo{author}{\bibfnamefont{S.}~\bibnamefont{Olmschenk}},
  \bibinfo{author}{\bibfnamefont{K.~C.} \bibnamefont{Younge}},
  \bibinfo{author}{\bibfnamefont{D.~L.} \bibnamefont{Moehring}},
  \bibinfo{author}{\bibfnamefont{D.~N.} \bibnamefont{Matsukevich}},
  \bibinfo{author}{\bibfnamefont{P.}~\bibnamefont{Maunz}}, \bibnamefont{and}
  \bibinfo{author}{\bibfnamefont{C.}~\bibnamefont{Monroe}},
  \bibinfo{journal}{Phys. Rev. A} \textbf{\bibinfo{volume}{76}},
  \bibinfo{pages}{052314} (\bibinfo{year}{2007}).

\bibitem[{\citenamefont{Hayes et~al.}(2007)\citenamefont{Hayes, Julienne, and
  Deutsch}}]{hayes:ultracold_exchange}
\bibinfo{author}{\bibfnamefont{D.}~\bibnamefont{Hayes}},
  \bibinfo{author}{\bibfnamefont{P.~S.} \bibnamefont{Julienne}},
  \bibnamefont{and} \bibinfo{author}{\bibfnamefont{I.~H.}
  \bibnamefont{Deutsch}}, \bibinfo{journal}{Phys. Rev. Lett.}
  \textbf{\bibinfo{volume}{98}}, \bibinfo{pages}{070501}
  (\bibinfo{year}{2007}).

\bibitem[{\citenamefont{Gorshkov et~al.}(2009)\citenamefont{Gorshkov, Rey,
  Daley, Boyd, Ye, Zoller, and Lukin}}]{gorshkov:alkaline_register}
\bibinfo{author}{\bibfnamefont{A.~V.} \bibnamefont{Gorshkov}},
  \bibinfo{author}{\bibfnamefont{A.~M.} \bibnamefont{Rey}},
  \bibinfo{author}{\bibfnamefont{A.~J.} \bibnamefont{Daley}},
  \bibinfo{author}{\bibfnamefont{M.~M.} \bibnamefont{Boyd}},
  \bibinfo{author}{\bibfnamefont{J.}~\bibnamefont{Ye}},
  \bibinfo{author}{\bibfnamefont{P.}~\bibnamefont{Zoller}}, \bibnamefont{and}
  \bibinfo{author}{\bibfnamefont{M.~D.} \bibnamefont{Lukin}},
  \bibinfo{journal}{Phys. Rev. Lett.} \textbf{\bibinfo{volume}{102}},
  \bibinfo{pages}{110503} (\bibinfo{year}{2009}).

\bibitem[{\citenamefont{Cowan}(1981)}]{cowan:spectra}
\bibinfo{author}{\bibfnamefont{R.~D.} \bibnamefont{Cowan}},
  \emph{\bibinfo{title}{The Theory of Atomic Structure and Spectra}}
  (\bibinfo{publisher}{University of California Press}, \bibinfo{year}{1981}).

\bibitem[{\citenamefont{{Bi\'{e}mont} et~al.}(2002)\citenamefont{{Bi\'{e}mont},
  Quinet, Dai, Zhankui, Zhiguo, Xu, and Svanberg}}]{biemont:YbRydberg}
\bibinfo{author}{\bibfnamefont{E.}~\bibnamefont{{Bi\'{e}mont}}},
  \bibinfo{author}{\bibfnamefont{P.}~\bibnamefont{Quinet}},
  \bibinfo{author}{\bibfnamefont{Z.}~\bibnamefont{Dai}},
  \bibinfo{author}{\bibfnamefont{J.}~\bibnamefont{Zhankui}},
  \bibinfo{author}{\bibfnamefont{Z.}~\bibnamefont{Zhiguo}},
  \bibinfo{author}{\bibfnamefont{H.}~\bibnamefont{Xu}}, \bibnamefont{and}
  \bibinfo{author}{\bibfnamefont{S.}~\bibnamefont{Svanberg}},
  \bibinfo{journal}{J. Phys. B} \textbf{\bibinfo{volume}{35}},
  \bibinfo{pages}{4743} (\bibinfo{year}{2002}).

\bibitem[{\citenamefont{Safronova et~al.}(2002)\citenamefont{Safronova,
  Johnson, Safronova, and Albritton}}]{safronova:yb-like}
\bibinfo{author}{\bibfnamefont{U.~I.} \bibnamefont{Safronova}},
  \bibinfo{author}{\bibfnamefont{W.~R.} \bibnamefont{Johnson}},
  \bibinfo{author}{\bibfnamefont{M.~S.} \bibnamefont{Safronova}},
  \bibnamefont{and} \bibinfo{author}{\bibfnamefont{J.~R.}
  \bibnamefont{Albritton}}, \bibinfo{journal}{Phys. Rev. A}
  \textbf{\bibinfo{volume}{66}}, \bibinfo{pages}{022507}
  (\bibinfo{year}{2002}).

\bibitem[{\citenamefont{Burshtein et~al.}(1974)\citenamefont{Burshtein,
  Verolainen, Komarovskii, Osherovich, and Penkin}}]{burshtein:delayed-coin}
\bibinfo{author}{\bibfnamefont{M.~L.} \bibnamefont{Burshtein}},
  \bibinfo{author}{\bibfnamefont{Y.~F.} \bibnamefont{Verolainen}},
  \bibinfo{author}{\bibfnamefont{V.~A.} \bibnamefont{Komarovskii}},
  \bibinfo{author}{\bibfnamefont{A.~L.} \bibnamefont{Osherovich}},
  \bibnamefont{and} \bibinfo{author}{\bibfnamefont{N.~P.}
  \bibnamefont{Penkin}}, \bibinfo{journal}{Opt. Spectrosc. (USSR)}
  \textbf{\bibinfo{volume}{37}}, \bibinfo{pages}{351} (\bibinfo{year}{1974}).

\bibitem[{\citenamefont{Blagoev et~al.}(1978)\citenamefont{Blagoev,
  Komarovskii, and Penkin}}]{blagoev:delayed-coin}
\bibinfo{author}{\bibfnamefont{K.~B.} \bibnamefont{Blagoev}},
  \bibinfo{author}{\bibfnamefont{V.~A.} \bibnamefont{Komarovskii}},
  \bibnamefont{and} \bibinfo{author}{\bibfnamefont{N.~P.}
  \bibnamefont{Penkin}}, \bibinfo{journal}{Opt. Spectrosc. (USSR)}
  \textbf{\bibinfo{volume}{45}}, \bibinfo{pages}{832} (\bibinfo{year}{1978}).

\bibitem[{\citenamefont{Andersen et~al.}(1975)\citenamefont{Andersen, Poulsen,
  Ramanujam, and {Petrakiev Petkov}}}]{andersen:rare-earths}
\bibinfo{author}{\bibfnamefont{T.}~\bibnamefont{Andersen}},
  \bibinfo{author}{\bibfnamefont{O.}~\bibnamefont{Poulsen}},
  \bibinfo{author}{\bibfnamefont{P.~S.} \bibnamefont{Ramanujam}},
  \bibnamefont{and} \bibinfo{author}{\bibfnamefont{A.}~\bibnamefont{{Petrakiev
  Petkov}}}, \bibinfo{journal}{Sol. Phys.} \textbf{\bibinfo{volume}{44}},
  \bibinfo{pages}{257} (\bibinfo{year}{1975}).

\bibitem[{\citenamefont{Rambow and Schearer}(1976)}]{rambow:hanle}
\bibinfo{author}{\bibfnamefont{F.~H.~K.} \bibnamefont{Rambow}}
  \bibnamefont{and} \bibinfo{author}{\bibfnamefont{L.~D.}
  \bibnamefont{Schearer}}, \bibinfo{journal}{Phys. Rev. A}
  \textbf{\bibinfo{volume}{14}}, \bibinfo{pages}{738} (\bibinfo{year}{1976}).

\bibitem[{\citenamefont{Lowe et~al.}(1993)\citenamefont{Lowe, Hannaford, and
  {M{\aa}rtensson--Pendrill}}}]{lowe:radiative}
\bibinfo{author}{\bibfnamefont{R.~M.} \bibnamefont{Lowe}},
  \bibinfo{author}{\bibfnamefont{P.}~\bibnamefont{Hannaford}},
  \bibnamefont{and} \bibinfo{author}{\bibfnamefont{A.-M.}
  \bibnamefont{{M{\aa}rtensson--Pendrill}}}, \bibinfo{journal}{Z. Phys. D}
  \textbf{\bibinfo{volume}{28}}, \bibinfo{pages}{283} (\bibinfo{year}{1993}).

\bibitem[{\citenamefont{Li et~al.}(1999)\citenamefont{Li, Svanberg, Quinet,
  Tordoir, and {Bi\'{e}mont}}}]{li:YbIIlifetime}
\bibinfo{author}{\bibfnamefont{Z.~S.} \bibnamefont{Li}},
  \bibinfo{author}{\bibfnamefont{S.}~\bibnamefont{Svanberg}},
  \bibinfo{author}{\bibfnamefont{P.}~\bibnamefont{Quinet}},
  \bibinfo{author}{\bibfnamefont{X.}~\bibnamefont{Tordoir}}, \bibnamefont{and}
  \bibinfo{author}{\bibfnamefont{E.}~\bibnamefont{{Bi\'{e}mont}}},
  \bibinfo{journal}{J. Phys. B} \textbf{\bibinfo{volume}{32}},
  \bibinfo{pages}{1731} (\bibinfo{year}{1999}).

\bibitem[{\citenamefont{Berends et~al.}(1993)\citenamefont{Berends, Pinnington,
  Guo, and Ji}}]{berends:beam-laser-YbII}
\bibinfo{author}{\bibfnamefont{R.~W.} \bibnamefont{Berends}},
  \bibinfo{author}{\bibfnamefont{E.~H.} \bibnamefont{Pinnington}},
  \bibinfo{author}{\bibfnamefont{B.}~\bibnamefont{Guo}}, \bibnamefont{and}
  \bibinfo{author}{\bibfnamefont{Q.}~\bibnamefont{Ji}}, \bibinfo{journal}{J.
  Phys. B} \textbf{\bibinfo{volume}{26}}, \bibinfo{pages}{L701}
  (\bibinfo{year}{1993}).

\bibitem[{\citenamefont{Pinnington et~al.}(1997)\citenamefont{Pinnington,
  Rieger, and Kernahan}}]{pinnington:lifetime}
\bibinfo{author}{\bibfnamefont{E.~H.} \bibnamefont{Pinnington}},
  \bibinfo{author}{\bibfnamefont{G.}~\bibnamefont{Rieger}}, \bibnamefont{and}
  \bibinfo{author}{\bibfnamefont{J.~A.} \bibnamefont{Kernahan}},
  \bibinfo{journal}{Phys. Rev. A} \textbf{\bibinfo{volume}{56}},
  \bibinfo{pages}{2421} (\bibinfo{year}{1997}).

\bibitem[{\citenamefont{Taylor et~al.}(1997)\citenamefont{Taylor, Roberts,
  {Gateva-Kostova}, Clarke, Barwood, Rowley, and Gill}}]{taylor:d52}
\bibinfo{author}{\bibfnamefont{P.}~\bibnamefont{Taylor}},
  \bibinfo{author}{\bibfnamefont{M.}~\bibnamefont{Roberts}},
  \bibinfo{author}{\bibfnamefont{S.~V.} \bibnamefont{{Gateva-Kostova}}},
  \bibinfo{author}{\bibfnamefont{R.~B.~M.} \bibnamefont{Clarke}},
  \bibinfo{author}{\bibfnamefont{G.~P.} \bibnamefont{Barwood}},
  \bibinfo{author}{\bibfnamefont{W.~R.~C.} \bibnamefont{Rowley}},
  \bibnamefont{and} \bibinfo{author}{\bibfnamefont{P.}~\bibnamefont{Gill}},
  \bibinfo{journal}{Phys. Rev. A} \textbf{\bibinfo{volume}{56}},
  \bibinfo{pages}{2699} (\bibinfo{year}{1997}).

\bibitem[{\citenamefont{Yu and Maleki}(2000)}]{yu:lifetime}
\bibinfo{author}{\bibfnamefont{N.}~\bibnamefont{Yu}} \bibnamefont{and}
  \bibinfo{author}{\bibfnamefont{L.}~\bibnamefont{Maleki}},
  \bibinfo{journal}{Phys. Rev. A} \textbf{\bibinfo{volume}{61}},
  \bibinfo{pages}{022507} (\bibinfo{year}{2000}).

\bibitem[{\citenamefont{Young et~al.}(1994)\citenamefont{Young, {Hill III},
  Sibener, Price, Tanner, Wieman, and Leone}}]{young:cs-single-photon}
\bibinfo{author}{\bibfnamefont{L.}~\bibnamefont{Young}},
  \bibinfo{author}{\bibfnamefont{W.~T.} \bibnamefont{{Hill III}}},
  \bibinfo{author}{\bibfnamefont{S.~J.} \bibnamefont{Sibener}},
  \bibinfo{author}{\bibfnamefont{S.~D.} \bibnamefont{Price}},
  \bibinfo{author}{\bibfnamefont{C.~E.} \bibnamefont{Tanner}},
  \bibinfo{author}{\bibfnamefont{C.~E.} \bibnamefont{Wieman}},
  \bibnamefont{and} \bibinfo{author}{\bibfnamefont{S.~R.} \bibnamefont{Leone}},
  \bibinfo{journal}{Phys. Rev. A} \textbf{\bibinfo{volume}{50}},
  \bibinfo{pages}{2174} (\bibinfo{year}{1994}).

\bibitem[{\citenamefont{Moehring et~al.}(2006)\citenamefont{Moehring, Blinov,
  Gidley, {Kohn, Jr.}, Madsen, Sanderson, Vallery, and
  Monroe}}]{moehring:lifetime}
\bibinfo{author}{\bibfnamefont{D.~L.} \bibnamefont{Moehring}},
  \bibinfo{author}{\bibfnamefont{B.~B.} \bibnamefont{Blinov}},
  \bibinfo{author}{\bibfnamefont{D.~W.} \bibnamefont{Gidley}},
  \bibinfo{author}{\bibfnamefont{R.~N.} \bibnamefont{{Kohn, Jr.}}},
  \bibinfo{author}{\bibfnamefont{M.~J.} \bibnamefont{Madsen}},
  \bibinfo{author}{\bibfnamefont{T.~D.} \bibnamefont{Sanderson}},
  \bibinfo{author}{\bibfnamefont{R.~S.} \bibnamefont{Vallery}},
  \bibnamefont{and} \bibinfo{author}{\bibfnamefont{C.}~\bibnamefont{Monroe}},
  \bibinfo{journal}{Phys. Rev. A} \textbf{\bibinfo{volume}{73}},
  \bibinfo{pages}{023413} (\bibinfo{year}{2006}).

\bibitem[{\citenamefont{Bell et~al.}(1991)\citenamefont{Bell, Gill, Klein,
  Levick, Tamm, and Schnier}}]{bell:four-level}
\bibinfo{author}{\bibfnamefont{A.~S.} \bibnamefont{Bell}},
  \bibinfo{author}{\bibfnamefont{P.}~\bibnamefont{Gill}},
  \bibinfo{author}{\bibfnamefont{H.~A.} \bibnamefont{Klein}},
  \bibinfo{author}{\bibfnamefont{A.~P.} \bibnamefont{Levick}},
  \bibinfo{author}{\bibfnamefont{C.}~\bibnamefont{Tamm}}, \bibnamefont{and}
  \bibinfo{author}{\bibfnamefont{D.}~\bibnamefont{Schnier}},
  \bibinfo{journal}{Phys. Rev. A} \textbf{\bibinfo{volume}{44}},
  \bibinfo{pages}{R20} (\bibinfo{year}{1991}).

\bibitem[{\citenamefont{Lehmitz et~al.}(1989)\citenamefont{Lehmitz,
  {Hattendorf--Ledwoch}, Blatt, and Harde}}]{lehmitz:pop-trap}
\bibinfo{author}{\bibfnamefont{H.}~\bibnamefont{Lehmitz}},
  \bibinfo{author}{\bibfnamefont{J.}~\bibnamefont{{Hattendorf--Ledwoch}}},
  \bibinfo{author}{\bibfnamefont{R.}~\bibnamefont{Blatt}}, \bibnamefont{and}
  \bibinfo{author}{\bibfnamefont{H.}~\bibnamefont{Harde}},
  \bibinfo{journal}{Phys. Rev. Lett.} \textbf{\bibinfo{volume}{62}},
  \bibinfo{pages}{2108} (\bibinfo{year}{1989}).

\bibitem[{not()}]{note:picoharp}
\bibinfo{note}{The TDC is a PicoQuant PicoHarp 300.}

\bibitem[{\citenamefont{Maunz et~al.}(2007)\citenamefont{Maunz, Moehring,
  Olmschenk, Younge, Matsukevich, and Monroe}}]{maunz:interference}
\bibinfo{author}{\bibfnamefont{P.}~\bibnamefont{Maunz}},
  \bibinfo{author}{\bibfnamefont{D.~L.} \bibnamefont{Moehring}},
  \bibinfo{author}{\bibfnamefont{S.}~\bibnamefont{Olmschenk}},
  \bibinfo{author}{\bibfnamefont{K.~C.} \bibnamefont{Younge}},
  \bibinfo{author}{\bibfnamefont{D.~N.} \bibnamefont{Matsukevich}},
  \bibnamefont{and} \bibinfo{author}{\bibfnamefont{C.}~\bibnamefont{Monroe}},
  \bibinfo{journal}{Nature Physics} \textbf{\bibinfo{volume}{3}},
  \bibinfo{pages}{538} (\bibinfo{year}{2007}).

\bibitem[{\citenamefont{Fawcett and Wilson}(1991)}]{fawcett:lifetimes}
\bibinfo{author}{\bibfnamefont{B.~C.} \bibnamefont{Fawcett}} \bibnamefont{and}
  \bibinfo{author}{\bibfnamefont{M.}~\bibnamefont{Wilson}},
  \bibinfo{journal}{Atomic Data and Nuclear Data Tables}
  \textbf{\bibinfo{volume}{47}}, \bibinfo{pages}{241} (\bibinfo{year}{1991}).

\bibitem[{\citenamefont{{Bi\'{e}mont} et~al.}(1998)\citenamefont{{Bi\'{e}mont},
  Dutrieux, Martin, and Quinet}}]{biemont:Ybcalc}
\bibinfo{author}{\bibfnamefont{E.}~\bibnamefont{{Bi\'{e}mont}}},
  \bibinfo{author}{\bibfnamefont{J.-F.} \bibnamefont{Dutrieux}},
  \bibinfo{author}{\bibfnamefont{I.}~\bibnamefont{Martin}}, \bibnamefont{and}
  \bibinfo{author}{\bibfnamefont{P.}~\bibnamefont{Quinet}},
  \bibinfo{journal}{J. Phys. B} \textbf{\bibinfo{volume}{31}},
  \bibinfo{pages}{3321} (\bibinfo{year}{1998}).

\bibitem[{\citenamefont{Safronova and
  Safronova}(2009)}]{safronova:3rd_order_Yb_calc}
\bibinfo{author}{\bibfnamefont{U.~I.} \bibnamefont{Safronova}}
  \bibnamefont{and} \bibinfo{author}{\bibfnamefont{M.~S.}
  \bibnamefont{Safronova}}, \bibinfo{journal}{Phys. Rev. A}
  \textbf{\bibinfo{volume}{79}}, \bibinfo{pages}{022512}
  (\bibinfo{year}{2009}).

\bibitem[{\citenamefont{DeVoe and Brewer}(1994)}]{devoe:eomlifetime}
\bibinfo{author}{\bibfnamefont{R.~G.} \bibnamefont{DeVoe}} \bibnamefont{and}
  \bibinfo{author}{\bibfnamefont{R.~G.} \bibnamefont{Brewer}},
  \bibinfo{journal}{Op. Lett.} \textbf{\bibinfo{volume}{19}},
  \bibinfo{pages}{1891} (\bibinfo{year}{1994}).

\bibitem[{\citenamefont{DeVoe and Brewer}(1996)}]{devoe:superradiance}
\bibinfo{author}{\bibfnamefont{R.~G.} \bibnamefont{DeVoe}} \bibnamefont{and}
  \bibinfo{author}{\bibfnamefont{R.~G.} \bibnamefont{Brewer}},
  \bibinfo{journal}{Phys. Rev. Lett.} \textbf{\bibinfo{volume}{76}},
  \bibinfo{pages}{2049} (\bibinfo{year}{1996}).

\bibitem[{\citenamefont{Haroche}(1976)}]{haroche:qbeats}
\bibinfo{author}{\bibfnamefont{S.}~\bibnamefont{Haroche}}, in
  \emph{\bibinfo{booktitle}{High-Resolution Laser Spectroscopy}}, edited by
  \bibinfo{editor}{\bibfnamefont{K.}~\bibnamefont{Shimoda}}
  (\bibinfo{publisher}{Springer-Verlag}, \bibinfo{year}{1976}),
  vol.~\bibinfo{volume}{13} of \emph{\bibinfo{series}{Topics in Applied
  Physics}}, p. \bibinfo{pages}{253}.

\bibitem[{\citenamefont{Silverman et~al.}(1978)\citenamefont{Silverman,
  Haroche, and Gross}}]{silverman:qbeats}
\bibinfo{author}{\bibfnamefont{M.~P.} \bibnamefont{Silverman}},
  \bibinfo{author}{\bibfnamefont{S.}~\bibnamefont{Haroche}}, \bibnamefont{and}
  \bibinfo{author}{\bibfnamefont{M.}~\bibnamefont{Gross}},
  \bibinfo{journal}{Phys. Rev. A} \textbf{\bibinfo{volume}{18}},
  \bibinfo{pages}{1507} (\bibinfo{year}{1978}).

\end{thebibliography}

\end{document}